\documentclass[runningheads]{llncs}

\usepackage[T1]{fontenc}
\usepackage{graphicx,verbatim}
\usepackage{amsmath}
\usepackage{amsfonts}
\usepackage{xfrac}
\usepackage{tabularx, booktabs}
\usepackage{multirow}
\usepackage{dsfont}
\usepackage{epsfig}
\usepackage{subcaption}
\usepackage[misc]{ifsym}

\newcolumntype{Y}{>{\centering\arraybackslash}X}

\newcommand{\FDG}{[$^{18}$F]FDG }

\newcommand{\imq}{\mathbf{q}}
\newcommand{\imA}{\mathbf{A}}

\newcommand{\imb}{\mathbf{b}}
\newcommand{\imm}{\mathbf{m}}
\newcommand{\imx}{\mathbf{x}}

\newcommand{\score}{\nabla_{\imx} \log p_t(\imx_t) }

\newcommand{\scorenn}{ \mathbf{s}_{\mathbf{\theta}}(\imx_t, t)}
\newcommand{\scorennempty}{\mathbf{s}_{\mathbf{\theta}}}
\newcommand{\scorenntwoempty}{\mathbf{s}_{\mathbf{\phi}}}

\newcommand{\tweedietheta}{\hat{\imx}_0(\imx_t, \scorennempty)}
\newcommand{\tweedieboth}{\hat{\imx}_0(\imx_t, \scorennempty + \scorenntwoempty)}

\begin{document}
\begin{titlepage}
\centering
\Large
\vspace*{\fill}
{\bfseries Supervised Diffusion-Model-Based PET Image Reconstruction\par}
\vspace{2\baselineskip}
This preprint is a de-anonymized version of a manuscript submitted for consideration at MICCAI 2025. This preprint has not undergone peer review or any post-submission improvements or corrections. 
\vspace*{\fill}
\end{titlepage}

\title{Supervised Diffusion-Model-Based PET Image Reconstruction}
%
%
\author{George Webber (\Letter)\inst{1} 
\and
Alexander Hammers\inst{1,2} \and
Andrew P. King\inst{1} \and
Andrew J. Reader\inst{1}}

\authorrunning{G. Webber et al. - Submitted Manuscript}
%
\institute{School of Biomedical Engineering and Imaging Sciences, King's College London, London, UK\\ %
\email{\{firstname\}.\{lastname\}@kcl.ac.uk} \and
Guy's \& St Thomas' PET Centre, St Thomas' Hospital, London, UK}
%
\maketitle
\begin{abstract}
Diffusion models (DMs) have recently been introduced as a regularizing prior for PET image reconstruction, integrating DMs trained on high-quality PET images with unsupervised schemes that condition on measured data. While these approaches have potential generalization advantages due to their independence from the scanner geometry and the injected activity level, they forgo the opportunity to explicitly model the interaction between the DM prior and noisy measurement data, potentially limiting reconstruction accuracy. To address this, we propose a supervised DM-based algorithm for PET reconstruction. Our method enforces the non-negativity of PET’s Poisson likelihood model and accommodates the wide intensity range of PET images. Through experiments on realistic brain PET phantoms, we demonstrate that our approach outperforms or matches state-of-the-art deep learning-based methods quantitatively across a range of dose levels.
We further conduct ablation studies to demonstrate the benefits of the proposed components in our model, as well as its dependence on training data, parameter count, and number of diffusion steps. Additionally, we show that our approach enables more accurate posterior sampling than unsupervised DM-based methods, suggesting improved uncertainty estimation. Finally, we extend our methodology to a practical approach for fully 3D PET and present example results from real \FDG brain PET data.


\end{abstract}

\section{Introduction}

Positron emission tomography (PET) is a widely used nuclear medicine technique for quantitatively imaging functional processes within the body \cite{shukla_positron_2006}. Clinicians and researchers rely on PET imaging to measure disease biomarkers, guide treatment decisions, and study physiological processes \cite{anand_clinical_2009}. 

Reducing the radioactive counts used for PET images is desirable to minimize radiation exposure and/or reduce scan times \cite{nievelstein_radiation_2012}. However, reconstructing PET images from limited radioactive counts is an ill-posed inverse problem due to the high-variance Poisson noise inherent in PET measurement data. State-of-the-art reconstruction algorithms address this issue by conditioning on image-based priors, which can be learned \cite{mehranian_model-based_2020,guazzo_learned_2021} or hand-crafted \cite{wang_penalized_2012}. These priors, combined with measured data, define a posterior distribution of possible reconstructed images \cite{filipovic_reconstruction_2021}. The posterior distribution reflects the uncertainty of the observed data, which is important in medical image reconstruction to quantify the reliability of image features and structures that may impact diagnosis \cite{filipovic_pet_2019}. However, most PET reconstruction methods only estimate the posterior mean \cite{reader_ai_2023}.

Recent work has leveraged diffusion models (DMs) \cite{sohl-dickstein_deep_2015,ho_denoising_2020} for PET reconstruction \cite{singh_score-based_2024,webber_likelihood-scheduled_2024,hu_patch-based_2024,webber_generative-model-based_2024,hu_unsupervised_2024,xie_joint_2024}. As the current state-of-the-art in generative image modeling \cite{dhariwal_diffusion_2021}, DM-based approaches offer two key advantages: 1) the potential to surpass the image quality of other deep-learned (DL) reconstruction approaches and 2) to fully model the posterior distribution of possible reconstructions from noisy data.

Previous approaches have integrated DMs trained on high-quality PET images with \textit{unsupervised} schemes that condition on measured data \cite{singh_score-based_2024,webber_likelihood-scheduled_2024,hu_patch-based_2024}. This enables these algorithms to remain agnostic to injected dose and scanner configuration, but they do not learn how the data consistency updates interact with the DM-based denoising updates, thereby compromising both quantitative accuracy and fidelity to the DM's learned manifold.

In this work, we propose and investigate an efficient \textit{supervised} conditioning framework tailored to PET reconstruction. Building on the DEFT (Doob’s h-transform Efficient Fine Tuning) framework proposed by Denker \textit{et al.} \cite{denker_deft_2024}, we pre-train a DM on high-quality PET images, which we then freeze while learning a measurement-conditional module with sinogram data. The design of our measurement-conditional module integrates solutions to several PET-specific challenges, such as the non-negativity of PET’s Poisson likelihood model and the wide intensity range of PET images.

The contributions of our paper are therefore:
\begin{itemize}
    \item We present the first supervised learning DM-based approach to PET image reconstruction.
    \item We propose and validate PET-specific modifications to existing DM-based frameworks for solving inverse problems.
    \item We match the performance of state-of-the-art supervised reconstruction approaches in our \textit{in-silico} experiments while also enabling posterior sampling.
    \item We demonstrate applicability to real-world clinical fully 3D \FDG brain PET data.
\end{itemize}

\section{Methodology}

\subsection{Preliminaries}

\subsubsection{PET reconstruction} PET reconstruction from measured sinograms is an inverse problem \cite{reader_deep_2021}. If $\imx$ is the true radiotracer distribution, $\imA$ is the system model (incorporating attenuation and normalization), and $ \imb $ models scatter and randoms components, the true mean $\imq$ of noisy measurements $\imm$ (e.g. a sinogram) may be modeled as
\begin{equation}
    \imq = \imA\imx + \imb \;, \; \imm \sim \text{Poisson}(\imq) \;.
\end{equation}

For an image estimate $\imx$, let the Poisson log-likelihood (PLL) of noisy measurements $\imm$ with respect to $\imx$ be $ L(\imx|\imm) := PLL(\max(\imx, 0) | \imm, \imA, \imb)$, and let $ p(\imx) $ be the prior probability density of $ \imx $.

\subsubsection{Diffusion models}
The central construct in a DM is a diffusion process mapping from high-quality images to unstructured Gaussian noise, often modeled as a stochastic differential equation (SDE) (over time $t$ from 0 to 1) \cite{song_score-based_2020}. This SDE defines transition probabilities $p_t$; the score vector $\score$ then represents the direction of removing artificial Gaussian noise from high-quality images. One may train a score network $\scorenn$ to estimate $\score$ by optimization with respect to the Denoising Score Matching (DSM) objective \cite{vincent_connection_2011}:
\begin{multline}\label{eq:DSM}
    \min_{\mathbf{\theta}} \big\{ \mathbb{E}_{t \sim U[0,1]} \mathbb{E}_{\imx_0 \sim p_0} \mathbb{E}_{\imx_t \sim p_t(\imx_t | \imx_0) } \bigl[ \| \scorenn - \nabla_{\imx} \log p_t(\imx_t|\imx_0) \|_2^2 \bigl]     \big\} \;.
\end{multline}
For image generation, we sample the initial image estimate $\imx_1$ as Gaussian noise, before iteratively denoising by approximating the reverse-time SDE using $\scorenn$. By interleaving denoising steps with unsupervised conditioning steps, or by training $\scorennempty$ with conditioning information, we may perform reconstruction.

These processes usually involve Tweedie's estimate $\tweedietheta = \frac{\imx_t + \sigma_t\scorenn}{\mu_t}$, which estimates the conditional expectation $\mathbb{E}(\imx_0 | \imx_t)$, given image iterate $\imx_t$ and trained score network $\scorenn$, where $\mu_t$ and $\sigma_t$ are terms analytically derived from $p_t$ (see \cite{singh_score-based_2024}). In this work, we use the variance-preserving SDE (see \cite{webber_likelihood-scheduled_2024}).

\begin{figure}[t]
    \centering
    \includegraphics[width=1\linewidth]{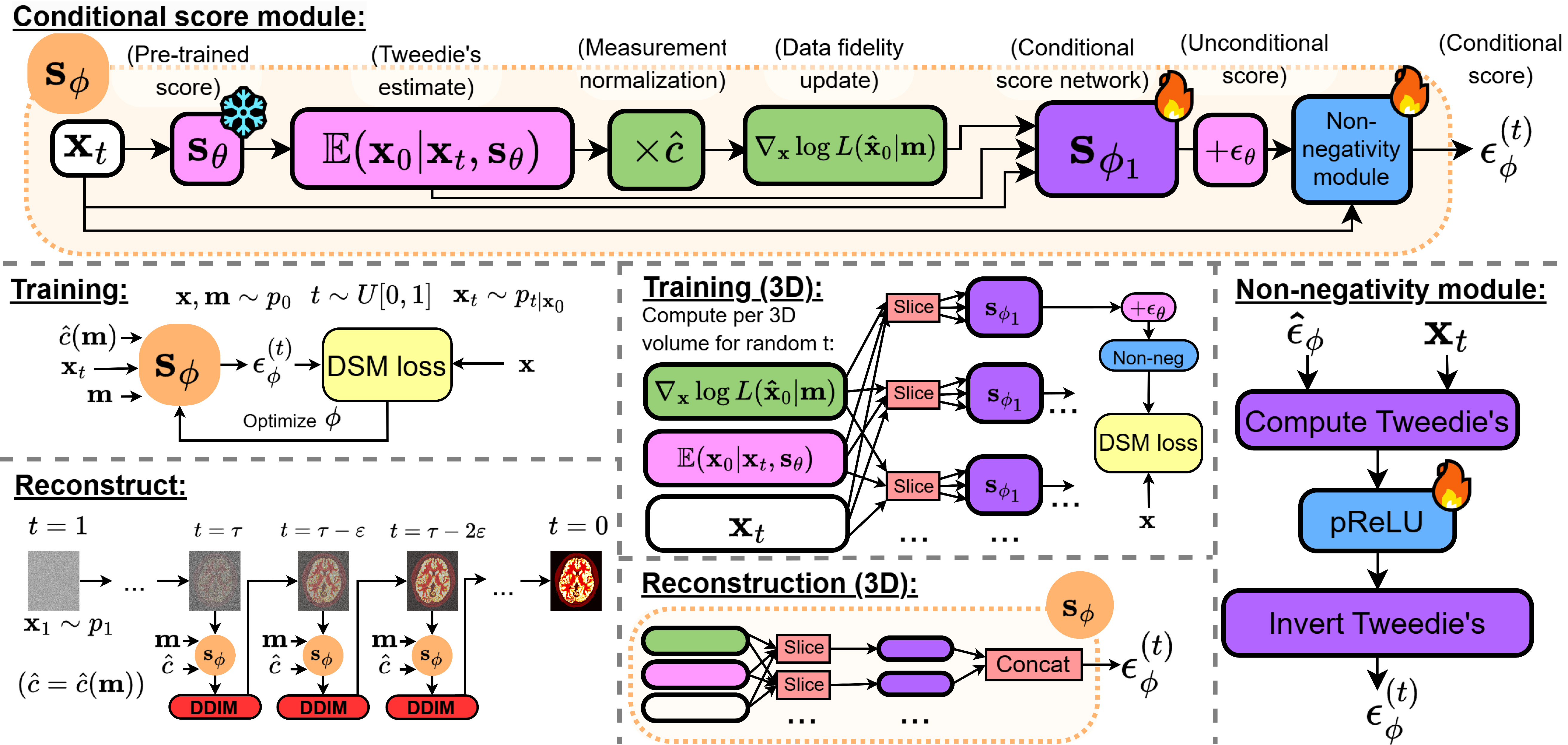}
    \caption{Methodology of proposed approach. The conditional score module $ \scorenntwoempty $ (top) learns to estimate a noisy image iterate $\imx_t$'s conditional score $\mathbf{\epsilon}_{\phi}^{(t)}$ (akin to a denoising vector).
    In $ \scorenntwoempty $, the unconditional Tweedie's estimate (from pre-trained $\scorennempty$) is measurement-normalized and used to compute a data fidelity update. Then $ \scorenntwoempty $ processes these terms, before adding $ \scorennempty $ and encouraging non-negativity of the conditional Tweedie's estimate, to output $\mathbf{\epsilon}_{\phi}^{(t)}$. Reconstruction uses $\mathbf{\epsilon}_{\phi}^{(t)}$ as a denoising direction to guide the DDIM scheme for reverse diffusion. In 3D, we precompute memory-intensive components of $\scorenntwoempty$ and train $\mathbf{s}_{\phi_{1}}$ as a 2.5D network on tranverse slices. We then reconstruct as in 2D, except with $\mathbf{\epsilon}_{\phi}^{(t)}$ formed by concatenating slice-wise outputs. Fire (ice) indicates trainable (frozen) modules.
    }
    \label{fig:methodology}
\end{figure}

\subsection{Approach}

Figure \ref{fig:methodology} presents an overview of our approach, PET-DEFT, named for its adaptation of the DEFT framework \cite{denker_deft_2024} to PET reconstruction.

To deal with PET images' wide dynamic range, we work in a normalized space for diffusion updates and un-normalize for data fidelity updates. Training on image data, we calculate scale factor $c = c(\imx)$; training or reconstructing from  measurements, we estimate $ c \approx \hat{c}(\imm)$, where $M$ is the $\mu$-map for dataset $i$ and $\imx_{\text{MLEM}_{10}}$ is the image estimate after 10 iterations of MLEM (Maximum-Likelihood Expectation-Maximization) \cite{shepp_maximum_1982}:
\begin{equation}
    c = c(\imx) := ||\imx||_1 / ||M||_0 \quad ; \quad  \hat{c} = \hat{c}(\imm) := || \mathds{1}_{M > 0}  \cdot \imx_{\text{MLEM}_{10}}||_1 / ||M||_0 \; .
\end{equation}

We firstly pre-train $\scorenn$ on $c$-normalized PET images (see \cite{singh_score-based_2024,webber_likelihood-scheduled_2024}). We then introduce paired datasets $(\imx^{(i)}, \imm^{(i)})_{i=1}^N$ from which to learn the conditional score network $\scorenntwoempty$. Given noisy image $\imx_t$, we calculate Tweedie's estimate using the pre-trained score $\tweedietheta$. This estimate is in normalized diffusion space; we transform it to calibrated measurement space via multiplication with $\hat{c}$. We then calculate a data fidelity update on this estimate as the gradient of the log-likelihood $\nabla_{\imx} \log L(\hat{c} \cdot \tweedietheta | \imm)$.

The noisy iterate $\imx_t$, Tweedie's estimate $\tweedietheta$, and the data fidelity update are then concatenated and inputted to the network $\scorenntwoempty$. The unconditional score $\scorenn$ is added, so that $\scorenntwoempty$ is trained to predict the residual.

Lastly, we apply a novel mechanism to encourage the non-negativity of our conditional Tweedie's estimate. This is motivated by a desire to keep the image iterate values $\imx_t$ positive so they are incorporated into later data fidelity updates (as the Poisson likelihood has the domain $\mathbb{R}^{\ge 0}$), thereby improving the conditional score estimation during test-time reconstruction. We calculate Tweedie's estimate from the conditional score $\tweedieboth$, apply a pReLU and invert Tweedie's estimate to get a score estimate corresponding to reduced negative values in $\tweedieboth$:
\begin{equation}
    \mathbf{\epsilon}_{\phi}^{(t)} =  \left( \mu_t \cdot \text{pReLU}(\tweedieboth) - \imx_t \right) / \sigma_t \; .
\end{equation}
We then freeze $\scorennempty$ and train the output $\mathbf{\epsilon}_{\phi}^{(t)}$ on the conditional DSM objective. To perform reconstruction, we use $\scorenntwoempty$ with the DDIM (Denoising Diffusion Implicit Model) \cite{song_denoising_2022} scheme (100 steps) for simulating the reverse diffusion SDE.

\subsubsection{Fully 3D reconstruction} Our approach extends efficiently to 3D, as we may decouple $\imA$, $\scorennempty$ and $\scorenntwoempty$. For a given 3D dataset, we first compute the pre-trained update components; we then train our conditional network on randomly-shuffled pre-computed slices of Tweedie's estimate, $\imx_t$ and the data fidelity update. We train a 2.5D network for $\scorennempty$ and $\mathbf{s}_{\phi_1}$, with $3\times$ as many input channels to accommodate information on 3 consecutive transverse slices (while retaining one output channel for predicting the score for the central slice). The 3D score estimate is then the concatenation of transverse slice-wise score estimates.

\begin{figure}[t]
    \centering
    \begin{subfigure}[b]{0.45\linewidth}
        \centering
        \includegraphics[width=\linewidth]{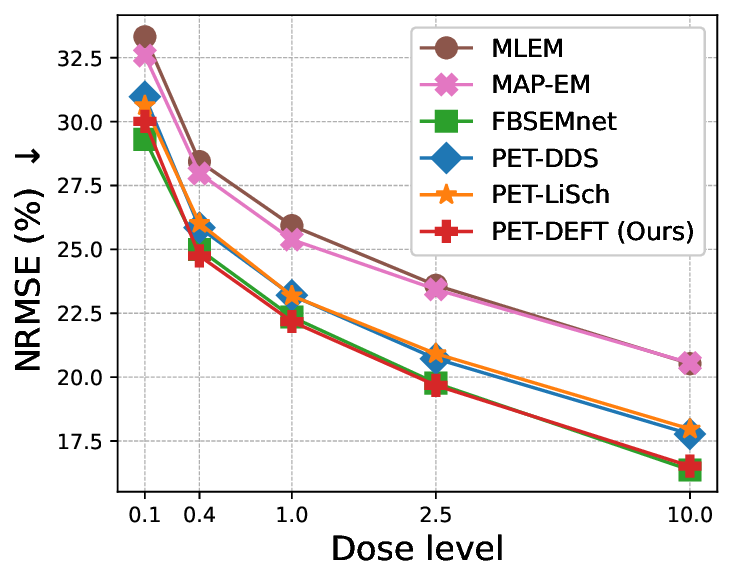}
        \caption{Normalized root mean square error}
    \end{subfigure}
    \hfill
    \begin{subfigure}[b]{0.45\linewidth}
        \centering
        \includegraphics[width=\linewidth]{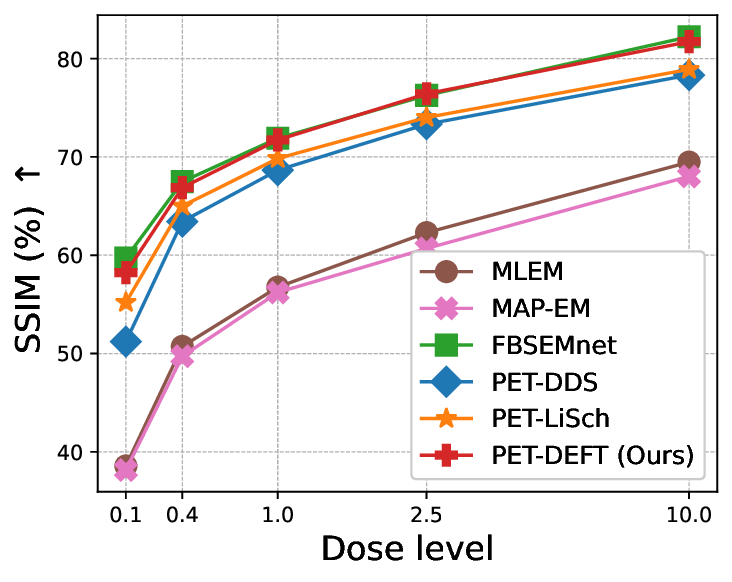}
        \caption{Structural similarity index measure}
    \end{subfigure}
    
    \caption{Quantitative accuracy of PET reconstruction algorithms at 5 dose levels.}
    \label{fig:quant_results}
\end{figure}

\section{Experiments}

\subsection{Experimental details}
\subsubsection{Dataset} 39 3D PET brain scans were simulated by segmenting real MR volumes and assigning realistic activity values to different tissues \cite{mehranian_model-based_2020}. Central 2D transverse slices were taken from these simulated scans to form the ground truth images used in this study. Measured data was generated by forward modeling each 2D image and applying Poisson noise to the resulting sinogram. Corrective factors were modeled as purely attenuation and contamination was modeled as a constant background of 30\% of simulated prompts. We define dose level 1.0 as when the average count rate was $1.7\times 10^6$.
For real data, we used 25 (2) previously acquired real 3D \FDG datasets for training (validation) \cite{mehranian_model-based_2020}. Real training images were MR-assisted reconstructions with the Bowsher prior \cite{schramm_evaluation_2018}.

\subsubsection{Implementation details} The forward model $\imA$ (common to all methods) was implemented with ParallelProj \cite{schramm_parallelprojopen-source_2024}, with dimensions approximating the Siemens Biograph mMR. 900 datasets were used for model training, 150 for validation and 10 for testing. Evaluations were made from three independent realizations of Poisson noisy sinograms. SSIM (structural similarity index measure) and LPIPS (learned perceptual image patch similarity) are perceptual metrics, while NRMSE (normalized root mean square error) is a distortion metric. NRMSE and SSIM values for DM-based methods were computed on the mean of 12 reconstructed samples (from the same noisy sinogram).
We implemented the model architecture in \cite{dhariwal_diffusion_2021} for unconditional and conditional score modules, varying the numbers of input channels as necessary. PyTorch \cite{paszke_pytorch_2019} and an Nvidia RTX 3090 GPU were used for experiments. Adam \cite{kingma_adam_2014} with learning rate $1 \times 10^{-4}$ ($8 \times 10^{-7}$) was used for 2D (3D) training. Supervised models were trained to minimize validation loss. Hyperparameters for MLEM, MAP-EM, PET-LiSch, and PET-DDS were tuned to minimize test-time NRMSE. In 2D, we set DDIM's stochasticity hyperparameter $\eta$ = 0.1 (following \cite{singh_score-based_2024,webber_likelihood-scheduled_2024}), while in 3D, we increased $\eta$ to 0.9, as this yielded results more visually consistent with the OSEM image.

\begin{figure}[t]
    \centering
    \epsfig{figure=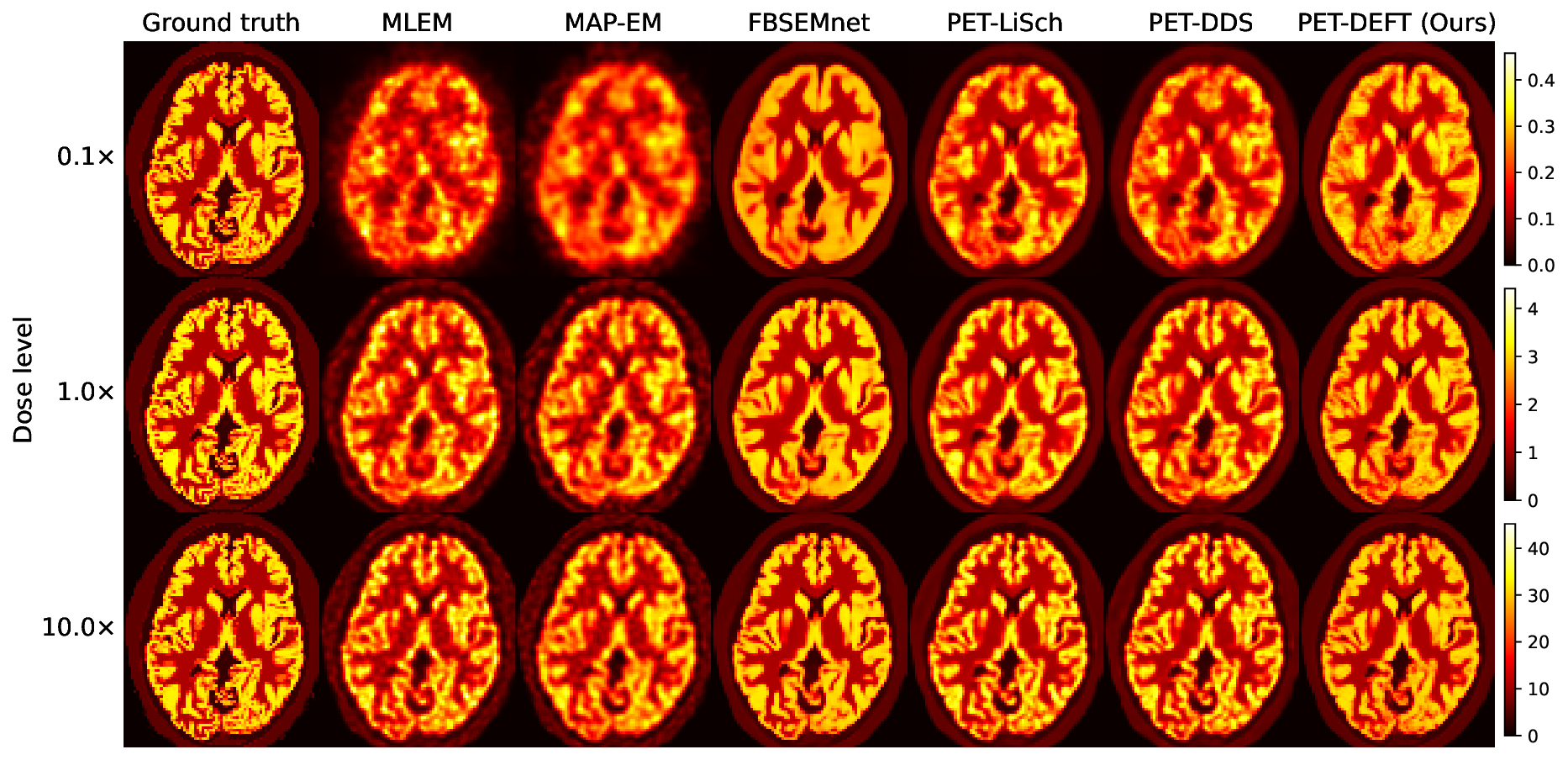, width=\linewidth}
    \caption{PET brain phantom reconstructions with increasing dose level.}
    \label{fig:quali_results}
\end{figure}

\begin{figure}[t]
    \centering
    \includegraphics[width=\linewidth]{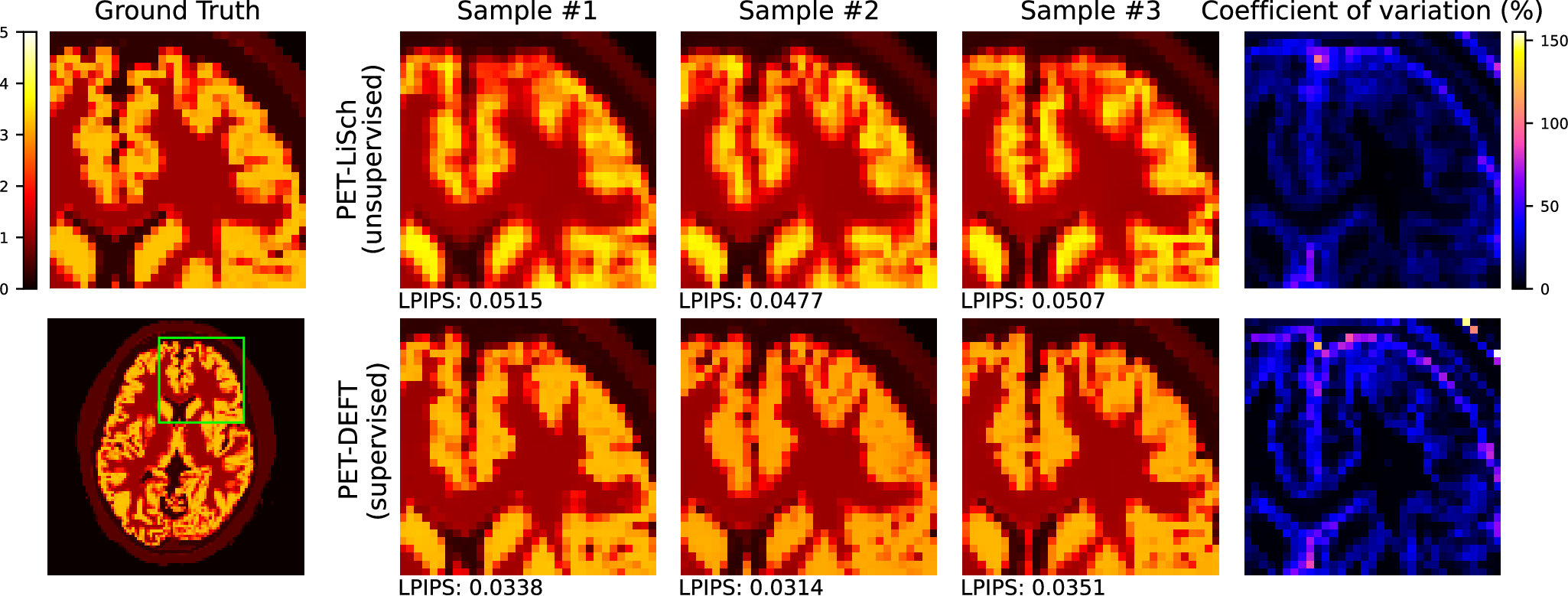}
    \caption{Example samples from unsupervised (top) and supervised (bottom) DM-based PET reconstruction (ours) at dose level 1.0.}
    \label{fig:posterior_sampling}
\end{figure}

\begin{figure}[t]
    \centering
    \begin{subfigure}[b]{0.45\linewidth}
        \centering
        \begin{minipage}[c][3cm]{\linewidth}
            \centering
            \begin{tabular}{c|c}
                \toprule
                Approach & NRMSE (\%) \\
                \midrule
                No normalization step & 30.0\\
                Measurement-normalized &  22.1\\
                \bottomrule
            \end{tabular}
        \end{minipage}
        \caption{Effect of measurement normalization}
    \end{subfigure}
    \hfill
    \begin{subfigure}[b]{0.45\linewidth}
        \centering
        \includegraphics[width=\linewidth]{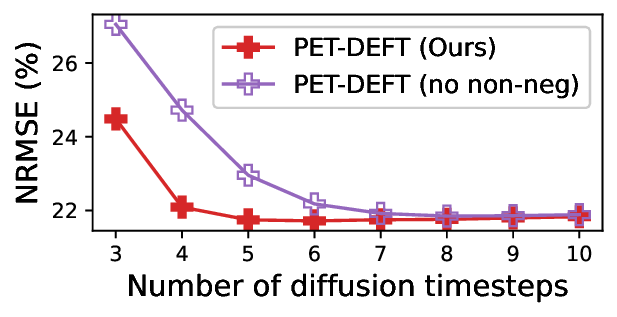}
        \caption{Effect of non-negativity module with few diffusion timesteps}
    \end{subfigure}

    \begin{subfigure}[b]{0.45\linewidth}
        \centering
        \includegraphics[width=\linewidth]{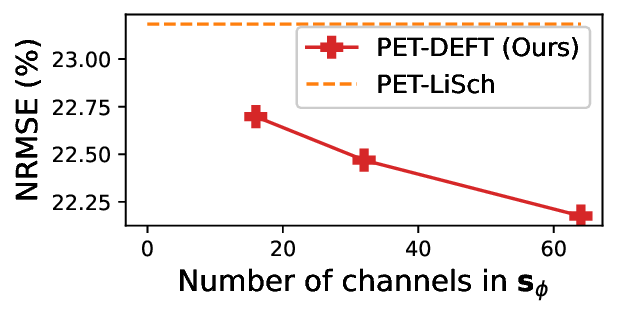}
        \caption{Number of parameters}
    \end{subfigure}
    \hfill
    \begin{subfigure}[b]{0.45\linewidth}
        \centering
        \includegraphics[width=\linewidth]{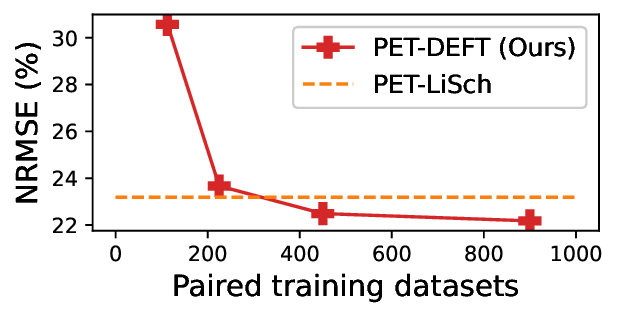}
        \caption{Number of training examples}
    \end{subfigure}

    \caption{Ablation studies into PET-DEFT's conditional score network (dose 1.0).}
    \label{fig:ablation_studies}
\end{figure}

\begin{figure}[t]
    \centering
    \includegraphics[width=1\linewidth]{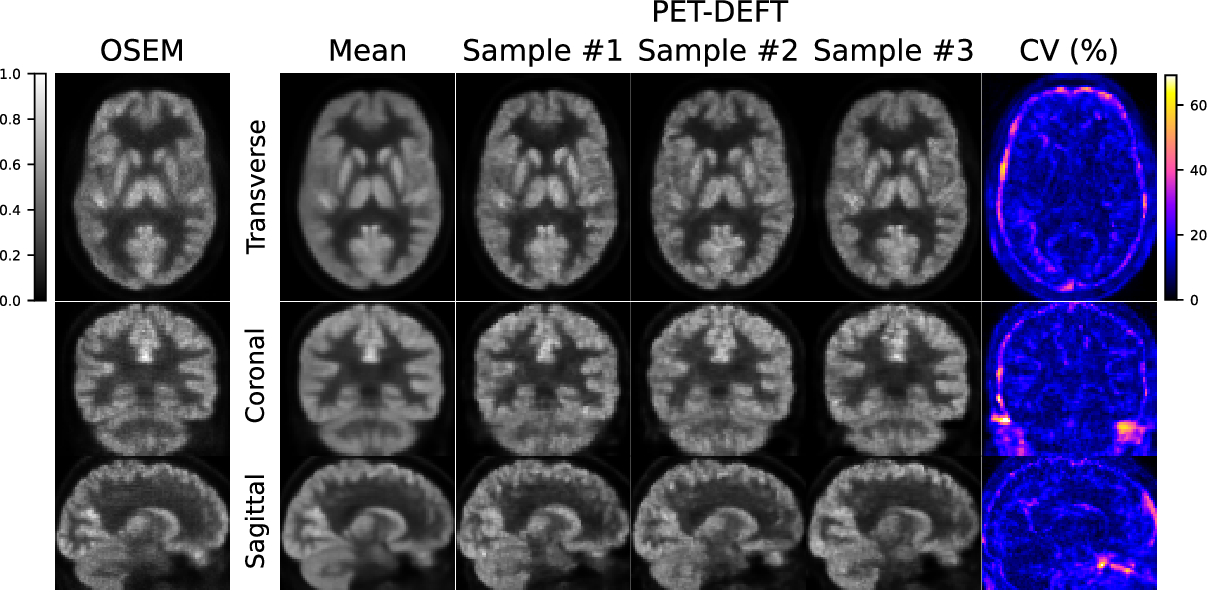}
    \caption{Slices from image reconstructed with PET-DEFT from 3D real data. Mean and coefficient of variation (CV) are calculated over 12 samples.}
    \label{fig:real_data_results}
\end{figure}

\subsection{Comparison against other methods}

MLEM is a clinical standard model-based algorithm \cite{shepp_maximum_1982}, while MAP-EM with a patch-based prior \cite{wang_penalized_2012} is a state-of-the-art model-based approach. PET-DDS \cite{singh_score-based_2024} and PET-LiSch \cite{webber_likelihood-scheduled_2024} are unsupervised DM-based PET reconstruction approaches. FBSEMnet is a supervised deep-learned unrolled iterative approach, representing the current state-of-the-art for deep-learned PET reconstruction \cite{mehranian_model-based_2020}.

\subsubsection{Quantitative analysis} Figure \ref{fig:quant_results} shows that our proposed method PET-DEFT performs on par with FBSEMnet, outperforming unconditional diffusion and model-based approaches across 5 dose levels. In general, the methods are grouped by performance into supervised DL, unsupervised DL and model-based methods.

\subsubsection{Qualitative analysis} Figure \ref{fig:quali_results} shows that our proposed approaches yield perceptually strong reconstruction accuracy at multiple dose levels. As counts decrease, PET-DEFT images resemble blurred scans (due to the averaging of multiple samples), retaining some fine-grained features, while FBSEMnet resembles a lower resolution blocky image (as it doesn't model the posterior distribution and is trained with an L2 loss).

\subsection{Posterior sampling and uncertainty}
Figure \ref{fig:posterior_sampling} shows single reconstruction samples from our supervised approach PET-DEFT and the unsupervised approach PET-LiSch. As is clear visually and also indicated by the LPIPS score, samples from PET-DEFT yield superior adherence to the real manifold of ground truth images.

\subsection{Ablation studies}

Figure \ref{fig:ablation_studies}(a) shows that the measurement normalization mechanism is essential for PET-DEFT's performance. Figure \ref{fig:ablation_studies}(b) shows that with the non-negativity module, we may perform as few as 5 reverse diffusion steps and retain our peak quantitative accuracy, compared to 8 steps without. 
In Figure \ref{fig:ablation_studies}(c) and (d), we show relative resilience of PET-DEFT to reducing the number of parameters or training sinogram-image pairs; the trend in Fig. \ref{fig:ablation_studies}(c) suggests possible further performance improvements with a larger conditioning network.

\subsection{Reconstruction from 3D Real \FDG data}

Compared to unrolled iterative methods such as FBSEMnet, our DM-based approach trains more efficiently (as the neural network and forward operator are accessed once each and decoupled for backpropagation), and also enables posterior sampling. We showcase these abilities by demonstrating reconstruction from real 3D \FDG sinogram data (using a commercial 24GB GPU) in Figure \ref{fig:real_data_results}.

\section{Conclusion}

In this paper, we propose novel and efficient methodology for supervised DM-based reconstruction of PET images. We introduce PET-specific mechanisms for training a conditional score network, including measurement normalization and a non-negativity module to address the wide dynamic range of PET images and the non-negativity of the Poisson likelihood. We found that our approach PET-DEFT outperforms unsupervised DM-based reconstruction and matches state-of-the-art PET reconstruction on experiments on simulated PET datasets. Our approach also enables more accurate posterior sampling for PET reconstruction. Lastly, we showed the feasibility and efficiency of our approach by reconstructing from real \FDG PET data, with an innovative training strategy that decouples the 3D PET physics operator from a 2.5D conditional score network.

\begin{credits}
\subsubsection{\ackname} GW acknowledges funding from the EPSRC CDT in Smart Medical Imaging [EP/S022104/1] and a GSK studentship. This work was also supported in part by EPSRC grant [EP/S032789/1]. The authors have applied a Creative Commons Attribution (CC BY) licence to any Accepted Author Manuscript version arising, in accordance with King’s College London’s Rights Retention policy.

\subsubsection{\discintname}
The authors have no competing interests to declare that are relevant to the content of this article.
\end{credits}


\newpage

\begin{thebibliography}{30}

\bibitem{shukla_positron_2006}
Shukla, A.K., Kumar, U.: Positron emission tomography: An overview. Journal of Medical Physics \textbf{31}(1), 13--21 (2006)

\bibitem{anand_clinical_2009}
Anand, S., Singh, H., Dash, A.: Clinical applications of PET and PET-CT. Medical Journal Armed Forces India \textbf{65}(4), 353--358 (2009)

\bibitem{nievelstein_radiation_2012}
Nievelstein, R.A.J., Quarles van Ufford, H.M.E., Kwee, T.C., Bierings, M.B., Ludwig, I., Beek, F.J.A., de Klerk, J.M.H., Mali, W.P.T.M., de Bruin, P.W., Geleijns, J.: Radiation exposure and mortality risk from CT and PET imaging of patients with malignant lymphoma. European Radiology \textbf{22}(9), 1946--1954 (2012)

\bibitem{mehranian_model-based_2020}
Mehranian, A., Reader, A.J.: Model-based deep learning PET image reconstruction using forward-backward splitting expectation-maximization. IEEE Transactions on Radiation and Plasma Medical Sciences \textbf{5}(1), 54--64 (2020)

\bibitem{guazzo_learned_2021}
Guazzo, A., Colarieti-Tosti, M.: Learned primal dual reconstruction for {PET}. Journal of Imaging \textbf{7}(12), ~248 (2021)

\bibitem{wang_penalized_2012}
Wang, G., Qi, J.: Penalized likelihood PET image reconstruction using patch-based edge-preserving regularization. IEEE Transactions on Medical Imaging \textbf{31}(12), 2194--2204 (2012)

\bibitem{filipovic_reconstruction_2021}
Filipovi\'c, M., Dautremer, T., Comtat, C., Stute, S., Barat, Ã.: Reconstruction, analysis and interpretation of posterior probability distributions of {PET} images, using the posterior bootstrap. Physics in Medicine and Biology \textbf{66}(12) (2021)

\bibitem{filipovic_pet_2019}
Filipovi\'c, M., Barat, E., Dautremer, T., Comtat, C., Stute, S.: {PET} {Reconstruction} of the {Posterior} {Image} {Probability}, {Including} {Multimodal} {Images}. IEEE Transactions on Medical Imaging \textbf{38}(7),  1643--1654 (2019)

\bibitem{reader_ai_2023}
Reader, A.J., Pan, B.: AI for PET image reconstruction. British Journal of Radiology \textbf{96}(1150), 20230292 (2023)

\bibitem{sohl-dickstein_deep_2015}
Sohl-Dickstein, J., Weiss, E., Maheswaranathan, N., Ganguli, S.: Deep unsupervised learning using nonequilibrium thermodynamics. In: Proceedings of the 32nd International Conference on Machine Learning, 2256--2265 (2015)

\bibitem{ho_denoising_2020}
Ho, J., Jain, A., Abbeel, P.: Denoising diffusion probabilistic models. Advances in Neural Information Processing Systems \textbf{33}, 6840--6851 (2020)

\bibitem{singh_score-based_2024}
Singh, I.R.D., Denker, A., Barbano, R., Kereta, Ž., Jin, B., Thielemans, K., Maass, P., Arridge, S.: Score-based generative models for PET image reconstruction. Machine Learning for Biomedical Imaging \textbf{2}(Generative Models), 547--585 (2024)

\bibitem{webber_likelihood-scheduled_2024}
Webber, G., Mizuno, Y., Howes, O.D., Hammers, A., King, A.P., Reader, A.J.: Likelihood-scheduled score-based generative modeling for fully 3D PET image reconstruction. arXiv preprint arXiv:2412.04339 (2024)

\bibitem{hu_patch-based_2024}
Hu, J., Song, B., Fessler, J.A., Shen, L.: Patch-{based} {diffusion} {models} {beat} {whole}-{image} {models} for {mismatched} {distribution} {inverse} {problems}. arXiv preprint arXiv:2410.11730 (2024)

\bibitem{webber_generative-model-based_2024}
Webber, G., Mizuno, Y., Howes, O.D., Hammers, A., King, A.P., Reader, A.J.: Generative-model-based fully 3D PET image reconstruction by conditional diffusion sampling. In: IEEE Nuclear Science Symposium and Medical Imaging Conference (NSS/MIC), 1--2 (2024)

\bibitem{hu_unsupervised_2024}
Hu, R., Wu, D., Tivnan, M., Guo, N., Yoon, S., Chen, Z., Wang, Y., Luo, J., Xue, S., Cui, J., Liu, H., Li, Q.: Unsupervised low-dose {PET} image reconstruction based on pre-trained denoising diffusion probabilistic prior. Journal of Nuclear Medicine \textbf{65}(supplement 2), 241109--241109 (2024)

\bibitem{xie_joint_2024}
Xie, T., Cui, Z.X., Luo, C., Wang, H., Liu, C., Zhang, Y., Wang, X., Zhu, Y., Chen, G., Liang, D., Jin, Q., Zhou, Y., Wang, H.: Joint diffusion: Mutual consistency-driven diffusion model for PET-MRI co-reconstruction. Physics in Medicine \& Biology \textbf{69}(15), 155019 (2024)

\bibitem{dhariwal_diffusion_2021}
Dhariwal, P., Nichol, A.: Diffusion {Models} {Beat} {GANs} on {Image} {Synthesis}. Advances in {Neural} {Information} {Processing} {Systems} \textbf{34}, 8780--8794 (2021)

\bibitem{denker_deft_2024}
Denker, A., Vargas, F., Padhy, S., Didi, K., Mathis, S.V., Barbano, R., Dutordoir, V., Mathieu, E., Komorowska, U.J., Lio, P.: {DEFT}: {Efficient} {fine}-tuning of diffusion models by {learning} the {generalised} $h$-transform. Advances in Neural Information Processing Systems \textbf{37}, 19636--19682 (2024)

\bibitem{reader_deep_2021}
Reader, A.J., Corda, G., Mehranian, A., Costa-Luis, C.d., Ellis, S., Schnabel, J.A.: Deep learning for PET image reconstruction. IEEE Transactions on Radiation and Plasma Medical Sciences \textbf{5}(1), 1--25 (2021)

\bibitem{song_score-based_2020}
Song, Y., Sohl-Dickstein, J., Kingma, D.P., Kumar, A., Ermon, S., Poole, B.: Score-based generative modeling through stochastic differential equations. In: International Conference on Learning Representations (ICLR) (2021)

\bibitem{vincent_connection_2011}
Vincent, P.: A connection between score matching and denoising autoencoders. Neural Computation \textbf{23}(7), 1661--1674 (2011)

\bibitem{shepp_maximum_1982}
Shepp, L.A., Vardi, Y.: Maximum likelihood reconstruction for emission tomography. IEEE Transactions on Medical Imaging \textbf{1}(2), 113--122 (1982)

\bibitem{song_denoising_2022}
Song, J., Meng, C., Ermon, S.: Denoising diffusion implicit models. arXiv preprint arXiv:2010.02502 (2022)

\bibitem{schramm_evaluation_2018}
Schramm, G., Holler, M., Rezaei, A., Vunckx, K., Knoll, F., Bredies, K., Boada, F., Nuyts, J.: Evaluation of parallel level sets and Bowsher’s method as segmentation-free anatomical priors for time-of-flight PET reconstruction. IEEE Transactions on Medical Imaging \textbf{37}(2), 590--603 (2018)

\bibitem{schramm_parallelprojopen-source_2024}
Schramm, G., Thielemans, K.: PARALLELPROJ—an open-source framework for fast calculation of projections in tomography. Frontiers in Nuclear Medicine \textbf{3} (2024)

\bibitem{paszke_pytorch_2019}
Paszke, A., Gross, S., Massa, F., Lerer, A., Bradbury, J., Chanan, G., Killeen, T., Lin, Z., Gimelshein, N., Antiga, L., Desmaison, A., Köpf, A., Yang, E., DeVito, Z., Raison, M., Tejani, A., Chilamkurthy, S., Steiner, B., Fang, L., Bai, J., Chintala, S.: PyTorch: An imperative style, high-performance deep learning library. Advances in Neural Information Processing Systems \textbf{33}, 8026--8037 (2019)

\bibitem{kingma_adam_2014}
Kingma, D.P., Ba, J.: Adam: A method for stochastic optimization. In: International Conference on Learning Representations (ICLR) (2015).
























\end{thebibliography}
\end{document}